\documentclass[9pt]{article}  \usepackage{times}
\usepackage{graphicx, textgreek}

\usepackage{color}

\usepackage[round,numbers,sort&compress]{natbib} 

\usepackage{amsmath}
\usepackage{nccmath}   
\usepackage{amssymb}
\usepackage{stackrel}
\usepackage{chemarrow}
\usepackage{graphicx}
\usepackage{textgreek}

\usepackage{booktabs}
\usepackage{dcolumn}
\usepackage{rotating}
\usepackage{multirow}

\renewcommand\appendix{\par
	\setcounter{section}{0}
	\setcounter{subsection}{0}
	\setcounter{figure}{0}
	\setcounter{table}{0}
	\renewcommand\thesection{Appendix \Alph{section}}
	\renewcommand\thefigure{\Alph{section}\arabic{figure}}
	\renewcommand\thetable{\Alph{section}\arabic{table}}
}

\begin{document}

	\twocolumn[{\LARGE \textbf{The thermodynamic soliton theory of the nervous impulse and possible medical implications\\*[0.2cm]}}
	{\large Thomas Heimburg$^\ast$\\*[0.1cm]
		{\small Niels Bohr Institute, University of Copenhagen, Blegdamsvej 17, 2100 Copenhagen \O, Denmark}\\*[-0.1cm]
		
		{\normalsize \textbf{ABSTRACT}\hspace{0.5cm} The textbook picture of nerve activity is that of a propagating voltage pulse driven by electrical currents through ion channel proteins, which are gated by changes in voltage, temperature, pressure or by drugs. All function is directly attributed to single molecules. We show that this leaves out many important thermodynamic couplings between different variables. A more recent alternative picture for the nerve pulse is of thermodynamic nature. It considers the nerve pulse as a soliton, i.e., a macroscopic excited region with properties that are influenced by thermodynamic variables including voltage, temperature, pressure and chemical potentials of membrane components. All thermodynamic variables are strictly coupled. We discuss the consequences for medical treatment in a view where one can compensate a maladjustment of one variable by adjusting another variable. For instance, one can explain why anesthesia can be counteracted by hydrostatic pressure and decrease in pH, suggest reasons why lithium over-dose may lead to tremor, and how tremor is related to alcohol intoxication. Lithium action as well as the effect of ethanol and the anesthetic ketamine in bipolar patients may fall in similar thermodynamic patterns. Such couplings remain obscure in a purely molecular picture. Other fields of application are the response of nerve activity to muscle stretching and the possibility of neural stimulation by ultrasound.
			\\*[0.3cm] }}
	\noindent\footnotesize{\textbf{Keywords:} Hodgkin-Huxley model, soliton theory, anesthesia, ion channels, thermodynamic couplings, lithium, tremor, nerve stretching, alcoholism\\*[0.1cm]}
	\noindent\footnotesize {$^{\ast}$corresponding author, theimbu@nbi.ku.dk. }\\
	\vspace{0.3cm}
	]

	\normalsize

\section{Introduction}
\label{introduction}

There exist two descriptions of nerve activity that are based on largely different concepts. The much-celebrated Hodgkin-Huxley model \cite[]{Hodgkin1952b} focuses on electrical phenomena such as voltage and currents. Functional control is due to the susceptibility of single protein and molecular receptors to changes in thermodynamic variables such as voltage, temperature, pressure and drugs. It is thus an intrinsically microscopic picture. Another more recent description of the nerve pulse is called the ``soliton-theory'' or the ``Heimburg-Jackson theory'' \cite[]{Heimburg2005c}. It is based on the assumption that the nerve membrane is a thermodynamic system and can be described using traditional thermodynamic concepts. Thus, the membrane responds as a whole to changes in thermodynamic variables.

The Hodgkin-Huxley (HH) model is a central pillar of biology and represents one of the founding myths of molecular biology. It describes the nerve signal as a voltage pulse. Its elements, the ion channel proteins, are tunable resistors that are specific for particular ions and respond to different stimuli. Evidence for their existence comes indirectly from patch clamp measurements \cite[]{Neher1976} that find electrical currents that are either on or off and from X-ray crystallography experiments (e.g., \cite{Doyle1998}) that revealed structures of numerous channel proteins.

It has become common to attribute large parts of biological activity to single molecules and reaction networks between them, which are called (reaction) pathways. Single molecules (especially protein receptors and channels) are typically of nanometer size. Focusing on such small scales makes it impossible to understand larger scale phenomena such as phase transitions, sound, waves and similar macroscopic phenomena that are likely to be of biological relevance. Such phenomena are called \emph{emergent} phenomena. It is virtually impossible to understand them quantitatively as the sum of molecular properties. Of even greater importance, this reservation also applies to the understanding of the thermodynamics of biological systems. Thermodynamics is our tool to understand the dependence of the state of a system on variables such as voltage, pressure, mechanical tension, temperature, chemical potentials and other intensive variables. It is clear from experiments that nerves respond to changes in these variables. Eschewing thermodynamics but in need of a mechanism to control the effects of these variables, molecular biology has introduced receptors, ion channels and other objects that respond to such changes. These include most notably ion channels that are gated by pressure and tension, temperature, voltage and drugs. While there is no doubt that the average state of single molecules depends on changes in the intensive variables, this does not imply that one can create a quantitatively reliable thermodynamics on the basis of single molecules. Important thermodynamics couplings, e.g., the Maxwell relations, cannot be explained on the level of single molecules. Therefore, any molecular attempt to understand the dependence of a macroscopic biological system on temperature, pressure, voltage or drugs must necessarily be incomplete and will most likely be wrong.

In this situation, we find it appropriate to turn to another theory, the ``soliton theory'', to describe the nerve pulse. It assumes that the nervous impulse is a sound pulse, i.e., an adiabatic density pulse during which the membrane changes area, thickness, and temperature. The soliton theory is of macroscopic thermodynamic nature, i.e., it is consistent with the laws of thermodynamics. The soliton (the technical term of a localized density pulse) is also of macroscopic nature, i.e., from a few millimeters to centimeters in length. Its description does not require objects on the molecular scale. From this perspective, ``ion channels'' arise as simple thermal fluctuations of the nerve membrane, i.e., they a consequence of macroscopic thermodynamics. The postulate is that the soliton and the nervous impulse are the same phenomenon.

We argue that the thermodynamic soliton theory is much better suited to make quantitative predictions. We will discuss what possible impact this could have for medical treatment. This includes anesthesia and some neurological diseases, and the effect of stretching on nerve activity which may be of interest to physiotherapy. Finally, we discuss the possibility of the stimulation of nerves by focussed ultrasound.


\section{The basic concepts}
\label{thebasicconcepts}


\subsection{Thermodynamics in neuroscience}
\label{thermodynamicsinneuroscience}

Thermodynamics is one of the basic pillars of classical physics. It is the theory of heat and work. The successes of thermodynamics are of such central importance that Einstein \cite{Einstein1949} said:

\begin{quote}
\emph{``It is the only physical theory of universal content concerning which I am convinced that, within the framework of the applicability of its basic concepts, it will never be overthrown (for the special attention of those who are skeptics on principle).''}
\end{quote}

Following Schr\"odinger \cite{Schroedinger1944}, we will assume that thermodynamics is applicable to biological systems, and that everything that does not agree with thermodynamics must be wrong.

The first law of thermodynamics states that changes in energy are equal to the work performed on the system and the heat added from the outside:

\begin{equation}
\label{eq:basic_1.1}
	dE=\underbrace{TdS}_{\mbox{heat}} + \underbrace{\left(... -\Pi dA+ ... +\Psi dq + ....+\sum_i \mu_i dn_i\right)}_{\mbox{work}} \;.
\end{equation}

Here, $TdS$ is the heat ($dS$ is the change in entropy and $T$ the temperature), and all other terms describe various kinds of mechanical, electrical or chemical work. When considering nerves, $TdS$ describes to the exchange of heat between the nerve and its surroundings. In a cylindrical axon, $-\Pi dA$ is the work performed by stretching or compressing the nerve membrane area, and $\Psi dq$ is the work of charging the membrane capacitor. Here, $\Psi$ is the electric potential and $q$ is the charge. The potential different between two points is $-\Delta \Psi = V$, i.e., a voltage. $\sum_i \mu_i dn_i$ represents chemical work that is associated to chemical reactions that might take place during the nerve pulse. This could, for instance, be a change in the protonation of the membrane surface, or other association reactions, but could also include the change of the population of different states of macromolecules, i.e., active and inactive states.


\subsection{The Hodgkin-Huxley model}
\label{thehodgkin-huxleymodel}

Hodgkin and Huxley \cite{Hodgkin1952b} assumed that the nerve axon is a cylinder composed of an electrolyte surrounded by a membrane. The axon is assumed to have three properties:

\begin{enumerate}
\item Its membrane is a conductor that selectively conducts sodium and potassium ions.

\item The membrane is also capacitor with a constant capacitance on the order of 1 $\mu$F\slash cm$^2$.

\item The cytosol inside and the medium outside of the cylinder is a conductor.

\end{enumerate}

The current through the membrane during the nerve pulse is given by
\begin{eqnarray}
\label{eq:basic_2.1}
I_m&=&\underbrace{C_m \frac{dV}{dt}}_{\mbox{capacitive current}} +\\ 
&& + \underbrace{g_K (V-E_K)+g_{Na} (V-E_{Na})+ ...}_{\mbox{ionic current}} \;,\nonumber
\end{eqnarray}
where $C_m$ is the membrane capacitance, $g_K$ is the conductance of the membrane for potassium, $g_{Na}$ is the conductance of the membrane for sodium and $V$ is the transmembrane voltage. $E_K$ and $E_{Na}$ are the Nernst potentials for potassium and sodium, respectively, that reflect the differences in the concentrations of potassium and sodium inside and outside of the axon. The Nernst potential is that voltage across a semipermeable membrane for which no current of the associated ion flows. Importantly, there exists no direct coupling between sodium and potassium fluxes, i.e., the potassium flux is independent of the sodium flux and vice versa.

\begin{figure}[htbp]
\centering
\includegraphics[width=7cm]{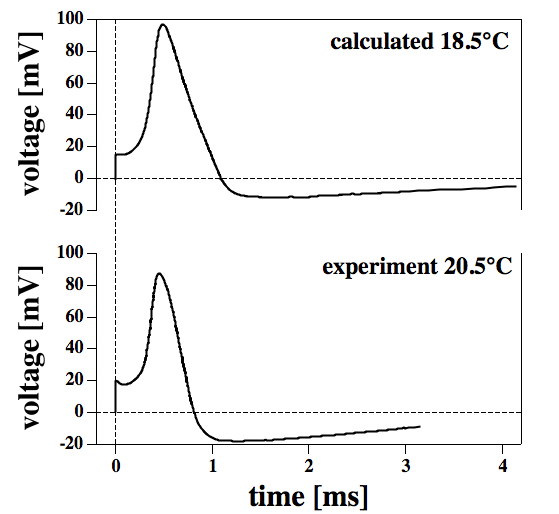}
\caption{Action potential in the squid giant axons at the indicated temperature. Top: Calculated from the HH-model. Bottom: Electrophysiological experiment. Adapted from \cite{Hodgkin1952b}.}
\label{hh_1952b_exp_vs_calc}
\end{figure}

Together with cable theory \cite[]{Hodgkin1946}, the differential equation for the propagation of an action potential yields
\begin{equation}
\label{eq:basic_2.2}
\frac{a}{2 R_i} \frac{\partial^2 V}{d x^2}=C_m \frac{dV}{dt} +g_K (V-E_K) + E_{Na} (V-E_{Na})+ ...  \;,
\end{equation}
where $a$ is the radius of the nerve and $R_i$ is the specific resistance of the cytosol inside of the nerve.
The solution of this differential equation, i.e. the time and space dependence of the voltage (solved numerically) is the so-called action potential representing the nerve pulse shown in Fig. \ref{hh_1952b_exp_vs_calc}. Both experimental and theoretical data are shown. The model describes the measured action potential in squid giant axons quite well.

The Hodgkin-Huxley model relies on electrical currents through the membrane and the associated voltage changes. The only term in eq. (\ref{eq:basic_2.1}) related to a change in the energy is the charging of the membrane capacitor, $\Psi dq$. The Hodgkin-Huxley model does not consider the stretching of the nerve and changes in the chemistry of the membrane. Temperature is also not contained explicitly. The flux of ions across the membrane is considered in analogy to the isothermal expansion of an ideal gas. The energy of an ideal gas does not depend on pressure and volume but only on temperature. Thus, the flux of particles does not lead to a change in energy under isothermal conditions as long as the total number of ions stays the same. However, since the flux of particles performs work by charging the membrane capacitor, one would expect an uptake of heat from the reservoir with exactly the same magnitude as the work performed on the capacitor \cite{Heimburg2021}. This is not contained in the Hodgkin-Huxley description but was later discussed by Howarth and collaborators \cite{Howarth1968} in an attempt to understand thermodynamic implications of the HH model.

One of the main problems of the Hodgkin-Huxley model lies in the description of the conductances, $g_K$ and $g_Na$, for which no theory exists. Therefore, for each temperature they have been fitted to experimental data, which requires 23 fitting parameters for each temperature (8 for the potassium conductance and 15 for the sodium conductance). Thus, the description of the action potential is a fit to the experimental data rather than a theory based on first principles. Given the simplicity of the shape of the action potential and the large number of adjustable parameters, the ability to reproduce the measured action potential should come as no surprise.

The conductances fitted by Hodgkin and Huxley were later attributed to ion channel proteins by Neher and Sakmann \cite{Neher1976}. Since a theoretical explanation of ion channel function is missing and not grounded in thermodynamics, the description of ion channel function by parametrization is the main reason for why a proper thermodynamic treatment of the Hodgkin-Huxley model is not possible. A second reason is the absence of any thermodynamic variable not related to voltage and currents. Additional channels sensitive to other variables will require the introduction of even more parameters. For instance, the Bostock model for myelinated axons described in \cite{Howells2012} contains 5 channels and 60--70 parameters.


\subsection{Pathways, metabolic cycles and thermodynamics}
\label{pathwaysmetaboliccyclesandthermodynamics}

More generally, the attribution of biological function to single molecules has led to an understanding according to which function can be understood by interactions of single molecules, ultimately leading to reaction networks and pathways. In some sense it is assumed that biology is a kind of clockwork in which chemical reactions are aligned in a sequential manner. The function of the cell is seen as the sum of the individual reactions. This way of looking at cells has one big problem: It is not consistent with thermodynamics.

The second law of thermodynamics states that the entropy of an isolated system will approach a maximum. This applies to the system as a whole but does not apply to parts of a system. Heimburg and Jackson \cite{Heimburg2007b} describe the case of two coupled gas pistons which are found in a position out of equilibrium. This example demonstrates that during equilibration the entropy of the two pistons together increases. However, this is not true for each individual piston since one of them decreases its entropy. If one considers only parts of a system, it is easy to make mistakes in the thermodynamics. While the entropy of the complete cell increases, this is not necessarily true for individual reactions or small parts of the cell. For instance, if within a cell two completely different reactions require calcium, these two reactions must be coupled since they compete for calcium. The same applies if reactions compete for drugs, protons, or for interfacial water. Thus, it is dangerous to regard individual reactions in a cell as distinct and separate from each other.

Thermodynamics recognizes many couplings between \linebreak thermodynamic variables of a complete system, most notably those expressed by the Maxwell relations. From eq. (\ref{eq:basic_1.1}) one can deduce
\begin{equation}
\label{eq:basic_3.1}
	\left(\frac{\partial V}{\partial T}\right)_S=	-\left(\frac{\partial S}{\partial p}\right)_V  \;,
\end{equation}
meaning that the temperature dependence of the volume in the absence of heat exchange is equal to the pressure-depen\-dence of the entropy at constant volume. Such relations are of immense practical importance because they show that all variables are coupled and they indicate how one can obtain relations such as $(\partial S/\partial p)_V$ from experiment.

Other Maxwell relations related to the molecular components of reactions are
\begin{equation}
\label{eq:basic_3.2}
	\left(\frac{\partial \mu_i}{\partial n_j}\right)_{n_i}=	\left(\frac{\partial \mu_j}{\partial n_i}\right)_{n_j}  \;,
\end{equation}
which implies that the dependence of the chemical potential $\mu_i$ of reagent $i$ (e.g., a protein) on the concentration of reagent $j$ (e.g., a drug or an ion) is equal to the dependence of the chemical potential $\mu_j$ of reagent $j$ on the concentration of reagent $i$. An example could be the coupling of the sodium concentration dependence of the chemical potential of the potassium channel and the potassium-dependence of the chemical potential of the sodium channel. This demonstrates that the chemical potentials of different moelcules depend on each other if one considers the complete ensemble (the cell or in the case of neuroscience, the complete nerve membrane).

Due to the focus on single molecules, membrane physiology postulates many molecular receptors and ion channels that are seemingly controlled by changes in intensive variables individually. Drug dependent receptors are controlled by the concentrations of drugs (e.g., \cite{Hille1992}), voltage-gated channels by the electrostatic potential (e.g., \cite{Bezanilla2005}), me\-chanosensitive channels by lateral tension (e.g., \cite{Syeda2016}), temperature sensitive channels by temperature (e.g., \cite{McKemy2002}) and acid-sensitive channels to pH changes (e.g., \cite{Gautam2010}). This seems necessary since it has been shown that nerve pulses can be excited by changes in voltage but also by temperature \cite[]{Kobatake1971, Shapiro2012} changes and by mechanical stimulation \cite[]{Tigerstedt1880}. However, if one class of ion channels is considered to be voltage-gated and another to be mechanosensitive, one loses the couplings contained in the Maxwell relations which are a property of an ensemble but not of a single molecule --- and thereby important information regarding the behavior of a biological system is lost. Therefore, a narrow molecular picture leads to errors in the understanding of the thermodynamics of the complete system. This is not to deny that proteins and small molecules are important. However, in order to obtain a complete picture one has to consider their chemical potentials within an ensemble rather than considering their individual molecular functions.


\subsection{Emergent and cooperative phenomena}
\label{emergentandcooperativephenomena}

There are numerous other phenomena which cannot be understood on the level of single molecules. This includes the propagation of sound (an entropy-conserving process), phase transitions such as the melting of ice or membranes (which are cooperative phenomena) or entropic effects such as freezing-point depression which will be shown to be important in understanding anesthesia.

\textbf{Waves:} Waves display a length and a time scale. Waves on water have length scales on the order of meters and velocities on the order of meters per second. Thus, they are about 10 orders of magnitude larger than the size of a water molecule. Waves can be surface waves or waves in the bulk medium that exist in water but also in oils, alcohols, or even in solids. Thus, the physics of waves depends on some macroscopic material constants but not on the chemistry of individual atoms, molecules or macromolecules. Waves are an emergent phenomena.

\textbf{Sound:} Sound is a particularly important kind of wave. When considering sound propagation in air, one usually focuses on periodic variations in density. The sound velocity in water is about 1500 m\slash s and about 150--200 m\slash s in the plane of liquid biological membranes. However, sound also is an adiabatic phenomenon where no heat is exchanged and the entropy is constant. Looking at eq. (\ref{eq:basic_1.1}), we see that at constant energy and entropy, all other variables couple. During sound propagation, density, temperature, polarization and even chemical equilibria can all change. Thus, an adiabatic wave is a wave with oscillations in all thermodynamic variables, i.e., it is not a purely mechanical phenomenon. In the soliton theory, the nerve pulse is considered as an adiabatic pulse reminiscent of sound.

\textbf{Phase transitions and melting:} Phase transitions are \linebreak changes of the state of a macroscopic system caused by a small change in an intensive variable. This could be the temperature, but also voltage, pressure, pH or the composition of the system (chemical potentials of the components). Large changes in the extensive variables occur over a relatively small interval of the intensive variables. The transitions in biomembranes display a width of about 10 degrees \cite[]{Heimburg2007a, Muzic2019}, although there was a recent report claiming the membrane melting in a human cell line (SH-SY5Ycells) can occur over the much smaller temperature interval of 1 degree \cite{Fedosejevs2022}.
Phase transitions and melting cannot be understood on the level of single molecules. They are cooperative phenomena that involve many molecules. The melting of ice cannot be understood by the investigation of single water molecules nor can the melting of biological membranes be understood by studying individual lipids and proteins.

\textbf{Entropic mixing and freezing point depression:}
In winter, when roads are kept free of ice by salt, one observes that salt can reduce the melting temperature of ice. This effect is called freezing-point depression, an effect first described by van't Hoff \cite[]{vantHoff1887}. It is explained by the entropy of ideal mixing of ions with liquid water and their insolubility in ice. It is an unspecific effect independent of the chemical nature of the solute. This purely entropic effect cannot be explained on the level of single molecules. As we show below, the effect of anesthetics on the melting of biomembranes can be explained by the same effect \cite[]{Heimburg2007c}. Anesthetics act as melting-point depressants in nerve membranes, independent of their chemical structure.


\begin{table*}[ht]
\begin{center}
\begin{tabular}{ |l||c|c| l |} 
 \hline
Nerve & velocity [m/s] &  length of pulse [mm] & source\\ 
\hline\hline
 lobster connective & 3.1 m/s & 3.1  & \cite{GonzalezPerez2014}\\ 
earth worm &  2.8-9.7  & 2.8-9.7 & \cite{GonzalezPerez2014}\\
 squid giant axon & 21.2  & 21.2 & \cite{Hodgkin1952b}\\ 
 frog sciatic nerve & 27.3 & 27.3 & \cite{Helmholtz1852}\\ 
cat peripheral nerve & up to 120 & up to 120 & \cite{Ritchie1982}\\
 human perineal nerve & 40 & 40 & \cite{Borg1984}\\
human median nerve & 65 & 65 & \cite{Buchthal1966}\\
 \hline
\end{tabular}
\end{center}
\caption{\label{table1}Length of the nerve pulse assuming a pulse duration of about 1 ms.}
\end{table*}

\subsection{Length scale of the nerve pulse}
\label{lengthscaleofthenervepulse}

Single molecule interactions play an important role on the molecular scale. However, macroscopic processes cannot simply be understood on a molecular scale. The applicability of a theory depends on the length and time scales it is made for. Thus, it would be important to consider the scales important for describing the function of neurons.

In contrast to many popular textbook drawings of neurons describing the action potential that suggest that the nerve pulse is a microscopic phenomenon, nerve pulses in fact are very large. Their length can easily be calculated by
\begin{equation}
\label{eq:basic_5.1}
	\Delta x= v\cdot \Delta t \;,
\end{equation}
where $\Delta x$ is the approximate length of the pulse, $v$ is its velocity and $\Delta t$ is the duration of the nerve pulse. The velocity $v$ in motor neurons can be up to 120 m\slash s \cite[]{Ritchie1982}, and in non-myelinated nerves it is on the order of 1--30 m\slash s (e.g., 3.1 m\slash s in the giant axons in lobster connectives \cite[]{GonzalezPerez2016}, and 21.2 m\slash s in giant axons of squid \cite[]{Hodgkin1952b}). The time scale of action potentials in most nerves is around 1 ms, much longer than molecular isomerization processes which are of the order of nanoseconds. This implies that the physical length of the nerve pulse is up to 12 cm in motor neurons. Some pulse dimensions are given in Table \ref{table1}. This is about 7 to 8 orders of magnitude larger than the scale of single molecules such as ion channel proteins with a diameter of about 5 nm. Nerve pulses are larger than most cells and even some small glia cells from the nervous system such as astrocytes that display a diameter of about 200 \textmu m \cite[]{Moshayedi2010}. Thus, the nerve pulse is a macroscopic phenomenon, and it seems plausible that it will require macroscopic physics in order to understand it.


\section{Thermodynamics of nerves}
\label{thermodynamicsofnerves}


\subsection{Melting transitions in nerve membranes}
\label{meltingtransitionsinnervemembranes}

Biological membranes can melt, i.e., they go from an ordered state at low temperature to a disordered state at high temperature. In these transitions, many properties of the membranes change, e.g., they absorb heat, they change their volume and area, and they become thinner. As a consequence, in the transition the heat capacity, the compressibility \cite[]{Heimburg2007a}, and the capacitance \cite[]{Heimburg2012} reach maxima. Such transitions have been observed in lung surfactant \cite[]{Ebel2001, BernardinodelaSerna2009}, and in \emph{E.coli} and \emph{b.subtilis} membranes, nerve membranes from pigs, sheep, and rats \cite{Muzic2019} and in various cancer cell membranes \cite[]{Hoejholt2019}. Recently, quite sharp transitions in human SH-SY5Ycells has been described \cite{Fedosejevs2022}. As we shall see, the existence of a melting transition slightly below body temperature is the prerequisite for the emergence of electromechanical solitons in nerve membranes.

\begin{figure*}[ht]
\centering
\includegraphics[width=310pt,height=149pt]{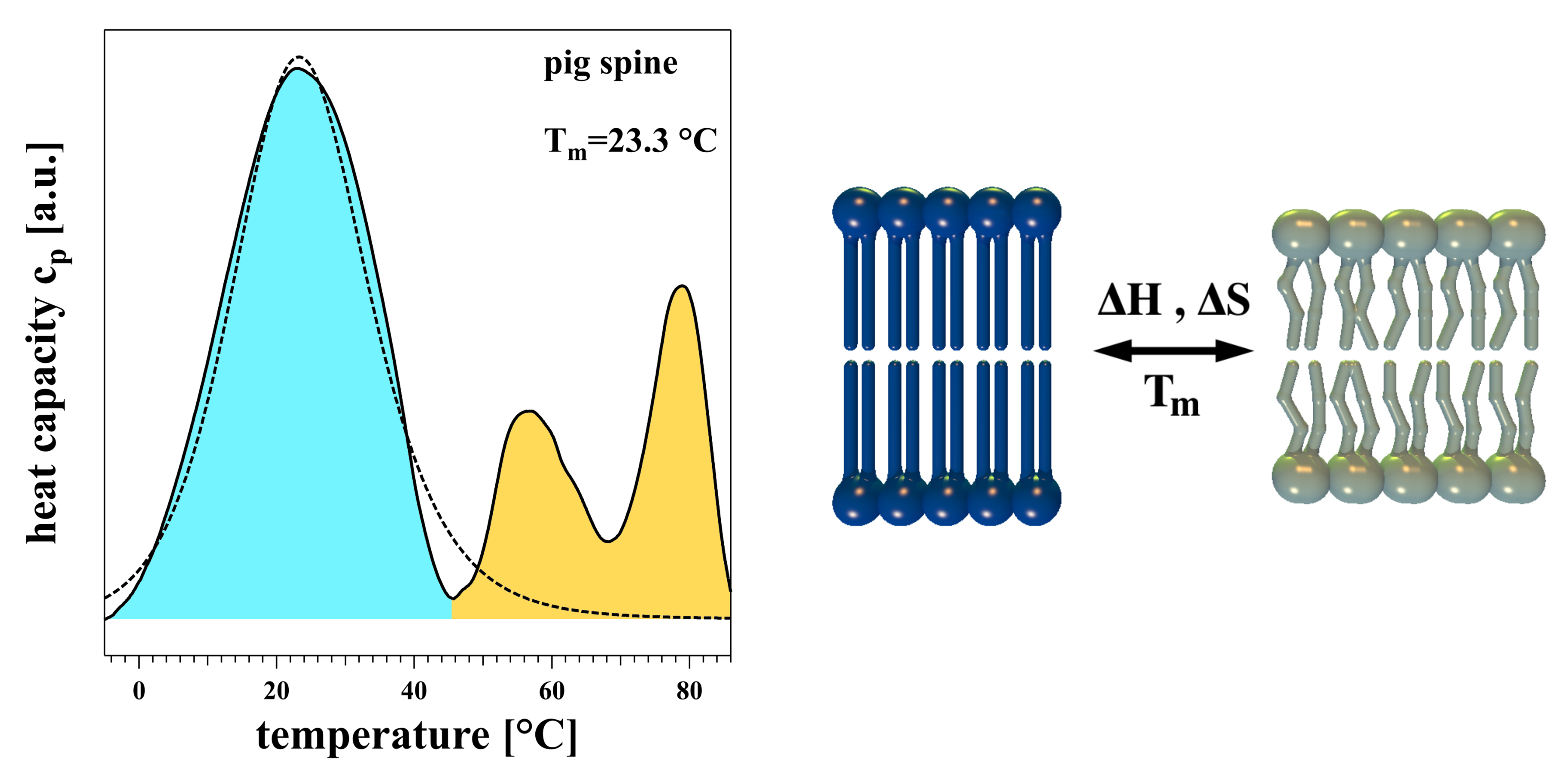}
\caption{Left: Heat capacity profile of the melting transition of membranes from pig spine. The blue shaded region is the melting of the membrane lipids at around 23.3$^\circ$C, i.e., below the body temperature of pigs at 39$^\circ$C. The yellow region represents the unfolding of membrane proteins above body temperature. Adapted from \cite{Muzic2019}. Right: Schematic drawing of the changes in the lipid membrane during a transition. Both, enthalpy $\Delta H$ and entropy $\Delta S$ change at the transition temperature $T_m$. }
\label{pig_spine_transition}
\end{figure*}

The transition temperature depends on lipid composition \cite[]{Heimburg2007a}, on pressure \cite[]{Ebel2001}, anesthetics concentration \cite[]{Heimburg2007c, Graesboll2014}, the nature of membrane proteins \cite[]{Heimburg1996b}, pH \cite{Trauble1976, Muzic2019, Fedosejevs2022} and other environmental conditions. For instance, anesthetic drugs lower the transition temperature but hydrostatic pressure and lowering of pH increases transition temperatures (see section \ref{anesthesia}).\\

\textbf{Adaptation:} It is important to note that the melting transitions are always below physiological temperature. In experiments where the growth temperature of bacteria is changed, the lipid composition adapts such that the melting temperature moves with growth temperatures. \emph{E. coli} bacteria shift the phase transition temperature with the growth temperature \cite[]{Wang2018}. Lipid composition adaptation to changes in growth conditions has been reported by various authors (see \cite{Heimburg2007a} for a discussion). E.g., the lipid composition in trout liver is different in summer and winter \cite[]{Hazel1979}. At cold temperature, the membrane contains a higher amount of unsaturated lipids with lower melting temperatures. Deep see bacteria adapt their lipid composition to hydrostatic pressure by favoring saturated lipids at high pressure \cite[]{DeLong1985}. Similar changes can be seen when bacteria grow in the presence of organic solvents \cite[]{Ingram1977}. Thus, it is to be expected that cell membranes adapt to the permanent presence of ethanol. For instance, chronic alcoholism in humans leads to decreases of polyunsaturated lipid concentrations \cite{Pita1997}. This attributes an important role to the metabolic control of the lipid composition and to the unsaturated fatty acids in particular.

The fastest possible adaptation is a pH-change that could compensate for sudden temperature changes, e.g., of frogs jumping from a sunny spot into a cold lake. In fact, it has been shown that the pH of frog muscle changes as a function of temperature. It is about 7.4 at 4$^\circ$C and about 6.9 at 37$^\circ$C, i.e., about half a pH unit between the two temperatures (\cite{Marjanovic1998}). \cite{Malan1976} report a pH-change of -0.0147 pH units per degree in frog muscle. Similar temperature-dependent changes in pH were also found in mouse soleus muscle \cite[]{Aickin1977}, in cardiac muscle \cite[]{Ellis1985} and in rat smooth muscle \cite[]{Dawson1985}. Generally, the pH is lower at higher temperatures, which implies that the transition temperature of the membranes is also higher. The lowering of pH from 9 to 5.5 in \emph{E.coli} membranes leads to an increase in transition temperature of about 7 degrees \cite[]{Muzic2019}.
Changes of pH in hibernating animals such as newts (-0.018 pH units per degree ) and salamander (-0.041 pH units per degree, \cite{Johnson1993}) were also of similar order (reviewed in \cite{Marjanovic1998}). The temperature-dependent pH changes have been attributed to a Na$^+$-H$^+$-exchange mechanism \cite[]{Marjanovic1998}.


\subsection{Elastic constants and sound in the transition}
\label{elasticconstantsandsoundinthetransition}

In the melting transition of membranes, the heat capacity is high. A small increase in temperature leads to a large uptake of heat. This is not the case away from the transition. Likewise, a small increase in pressure leads to a large decrease in volume because the volume in the gel membrane is smaller than that of the fluid membrane. This implies that the membrane displays a high compressibility in the melting transition. The area compressibility is similarly high in the vicinity of the transition \cite[]{Heimburg1998}. Thus, biological membranes exists close to a state where the membrane is softer and more compressible. At the heat capacity maximum, membranes also display maxima in volume and area compressibility. Interestingly, increase in pressure will also lead to a release of heat, and increase in temperature to an increase in area, i.e., there exist significant cross-couplings between non-conjugated variables.

It has been shown that one can calculate the changes in isothermal volume and area compressibility from the heat capacity \cite[]{Heimburg1998, Ebel2001}:
\begin{eqnarray}
\label{eq:nerves_2.1}
\Delta K_T^V &=& \frac{\gamma_V^2 T}{\left<V\right>}\Delta c_p \qquad\mbox{volume compressibility}\nonumber\\
\Delta K_T^A &=& \frac{\gamma_A^2 T}{\left<A\right>}\Delta c_p \qquad\mbox{area compressibility}
\end{eqnarray}
i.e., one can calculate the isothermal volume and area compressibility from the heat capacity profiles. The value of the coefficients $\gamma_V$ and $\gamma_A$ are fairly similar for a large range of artificial and biological membranes ($\gamma_V=7.8\cdot 10^{-10}$m$^2/$N and $\gamma_A=0.89$ m\slash N). This is very convenient because it allows for the estimation of both volume and area compressibility from the heat capacity even for membranes where neither composition nor phase behavior is known, e.g., for biological membranes \cite[]{Muzic2019}.

The sound velocity $c$ in membranes is a function of the adiabatic area compressibility $\kappa_S^A$
\begin{equation}
\label{eq:nerves_2.2}
c^2=\frac{1}{\kappa_S^A\cdot \rho^A} \;,
\end{equation}
where $\rho^A$ is the lateral mass density of the membrane. $\kappa_S^A$ is simply related to the isothermal compressibility $\kappa_T^A$ \cite{Heimburg1998}. For processes slow compared with the relaxation times in membranes, the adiabatic compressibility is equal to the isothermal compressibility \cite[]{Mosgaard2013a}. This implies that in the melting transition, the sound velocity displays a minimum, which is also what is found experimentally \cite[]{Halstenberg1998, Schrader2002}. This fact leads to the emergence of solitons in biological membranes and nerves \cite[]{Heimburg2005c}.


\subsection{Solitons}
\label{solitons}

Solitons are a special kind of waves that can emerge when the elastic constants are non-linear and dispersion (frequency dependence of the sound velocity) is present. Instead of obtaining a sinusoidal wave one finds a single pulse. The most commonly known soliton is that of a pulse-like wave in shallow water first described by John Scott Russell in 1834. In lipid membranes, the non-linearity is obtained because the sound velocity is density dependent \cite[]{Heimburg2005c}, and dispersion arises because the sound velocity is frequency-dependent \cite[]{Mitaku1982}.

The wave equation for sound in one dimension is given by
\begin{equation}
\label{eq:nerves_3.1}
\frac{\partial^2 \Delta \rho}{\partial t^2}=\frac{\partial}{\partial x}\left[c^2\frac{\partial \Delta \rho}{\partial x}\right]\;.
\end{equation} 

\begin{figure*}[htbp]
\centering
\includegraphics[width=12cm]{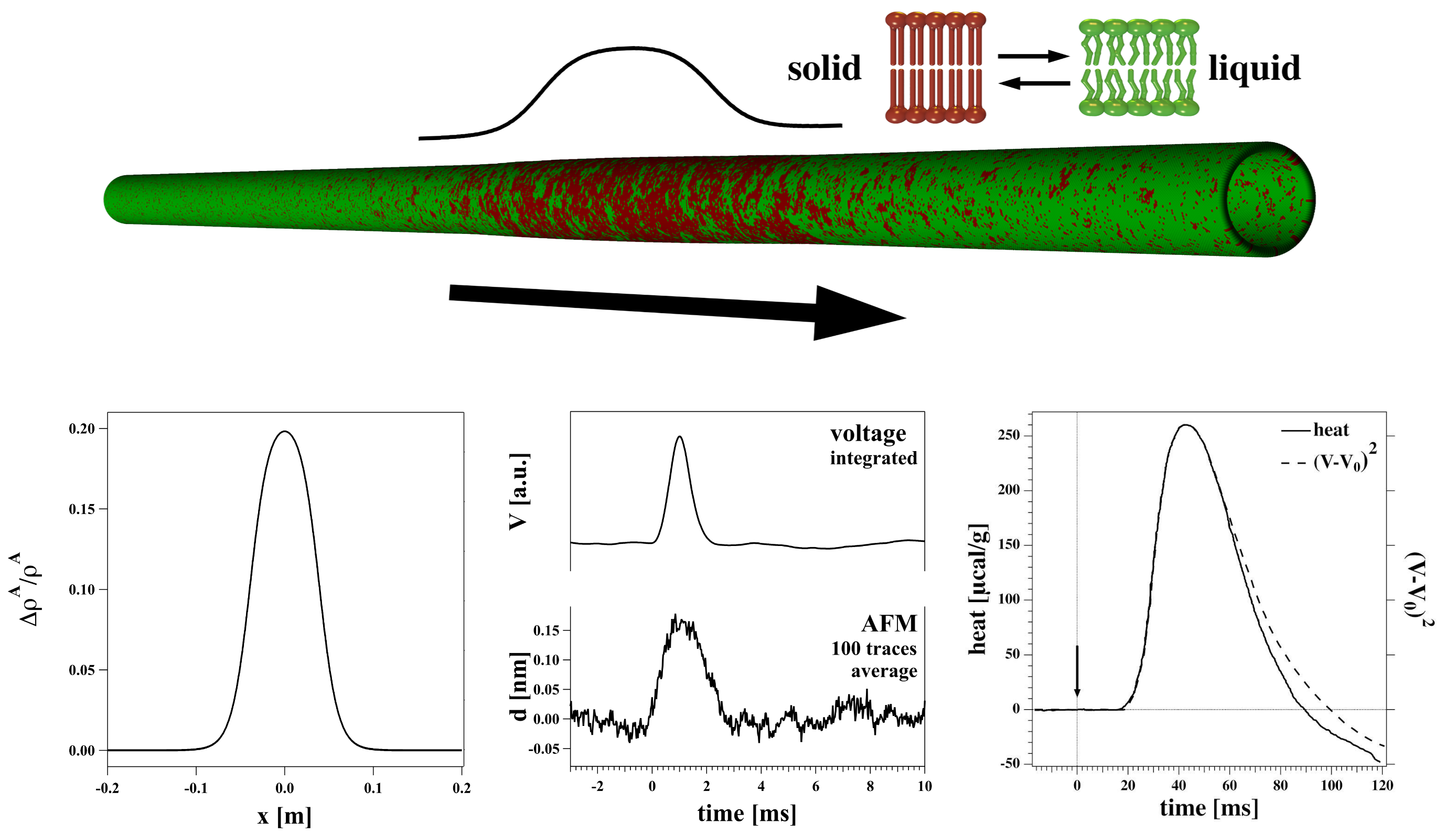}
\caption{Schematic view of the electromechanical soliton in a cylinder of a membrane consisting of a region in which the lipids are more ordered than outside of the pulse. Note that in real nerves the pulse is much longer than shown in this figure. Bottom, left: Soliton calculated for a pulse in a motoneuron. Adapted from \cite{Heimburg2005c}. Bottom, center: During the action potential in lobster connectives, the voltage pulses is accompanied by a change in the nerve thickness of the order of 0.15--1.0 nm. Adapted from \cite{GonzalezPerez2016}. Bottom, right: During the action potential in garfish olfactory nerves one finds a transient release of heat that is proportional to the square of the voltage. Adapted from \cite{Ritchie1985}. Note that the action potentials in lobster connective assume a different voltage profile as those in the squid axon shown in Fig. \ref{hh_1952b_exp_vs_calc}.}
\label{soliton_gonzalezperez2016b}
\end{figure*}

Close to melting transitions, the sound velocity is a nonlinear function of density. In second order approximation it is given by $c^2=c_0^2 + p\Delta \rho +q\Delta \rho^2+...$ Thus Heimburg and Jackson \cite{Heimburg2005c} were led to write the wave equation for a membrane close to the phase transition as
\begin{equation}
\label{eq:nerves_3.2}
\frac{\partial^2 \Delta \rho}{\partial t^2}=\frac{\partial}{\partial x}\left[\left(c_0^2+p\Delta \rho + q\Delta \rho^2\right)\frac{\partial \Delta \rho}{\partial x}\right]-\frac{\partial^4 \Delta \rho}{\partial x^4} \;.
\end{equation}
Here, $c_0$ is the sound velocity of a membrane in the fluid phase, $p$ and $q$ describe the shape of the melting transition as a function of density taken from experimental data. The last term in eq. (\ref{eq:nerves_3.2}) has been introduced to describe dispersion, and $h$ is a dispersion coefficient \cite[]{Heimburg2005c}. A solution of the above differential equation is given in Fig. \ref{soliton_gonzalezperez2016b}, bottom left. The coordinate x denotes the position along the axon at a fixed time, and the y-axis denotes the change in thickness of the nerve membrane. At the maximum of the pulse, the density change corresponds approximately to the difference between the thickness of the fluid and the solid state of the membrane.

Fig. \ref{soliton_gonzalezperez2016b} (top) schematically shows that during the propagation of the nerve pulse the membrane becomes thicker and the lipids more solid (shown in red). Therefore, the nerve pulse is also accompanied by a transient release of heat. Both the increase in thickness \cite[]{Iwasa1980a, Iwasa1980b, Tasaki1989, Kim2007, GonzalezPerez2016} and the transient release of heat \cite[]{Abbott1958, Howarth1968, Howarth1975, Ritchie1985, Heimburg2021} have been observed in experiments (see Fig. \ref{soliton_gonzalezperez2016b} (bottom, center) and \ref{soliton_gonzalezperez2016b} (bottom, right)). Thus, the nerve pulse displays not only voltage changes but also changes in thickness and temperature. Pulse-like patterns in both density and voltage have been shown to exist on lipid monolayers close to transitions in the complete absence of proteins \cite[]{Griesbauer2012a, Griesbauer2012b, Shrivastava2014}. In these studies it has been stressed that all thermodynamic variables couple \cite[]{Schneider2021}. Further, because the soliton is a region of solid phase in a fluid membrane, during pulse propagation the nerve will contract due to the lower area density of the gel phase. This has also been observed experimentally \cite[]{Wilke1912b, Tasaki1989}.

\textbf{Nerve pulse excitation:} The excitation of the soliton requires a sudden change in any variable that can move the melting transition through its phase transition. The stimulation threshold is proportional to the distance of physiological temperature from the membrane melting temperature and corresponds to a critical stimulation free energy \cite[]{Wang2018}. This can be voltage pulses (as commonly used in electrophysiology), mechanical stimulation, pH drops, temperature decrease \cite[]{Kobatake1971} and other sudden stimuli that increase the phase transition temperature. Nerve excitation can be inhibited by anesthetics \cite[]{Wang2018} and potentially by temperature increases, increases in pH, and other changes in variables that lower phase transition temperatures.
None of these changes are even considered --- let alone predicted --- in the Hodgkin-Huxley model. However, they are both predicted by \cite{Heimburg2005c} and these predictions have been confirmed.


\section{Anesthesia}
\label{anesthesia}

\begin{figure}[htbp]
\centering
\includegraphics[width=8.5cm]{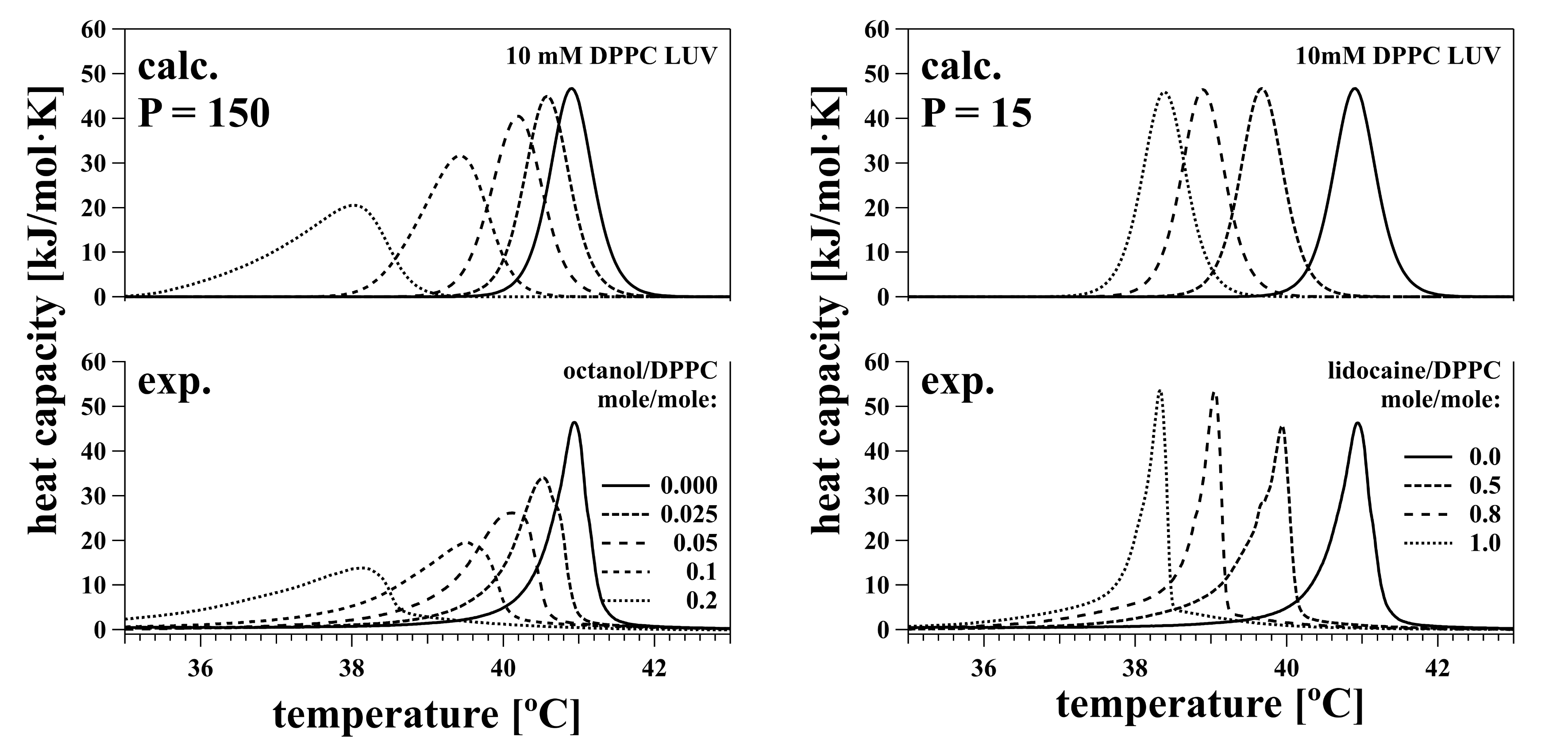}
\caption{Influence of anesthetics on the melting behavior of DPPC membranes. Left: The general anesthetic octanol. Right: The local anesthetic lidocaine. The bottom traces are experimental data, the top traces are profiles calculated assuming that van't Hoff's freezing-point depression law is valid. Both, general and local anesthetics shift melting transitions towards lower temperatures. The theoretical treatment describes the experimental profiles quite well. Adapted from \cite{Graesboll2014}.}
\label{graesboll2014}
\end{figure}

General anesthetics display a peculiar property: Their oil-water partition coefficient $P$ is exactly proportional to the inverse of the critical dose {[ED]}$_{50}$ that anesthetizes 50\% of the individuals \cite[]{Overton1991, Heimburg2007a, Heimburg2007c}. This relationship can be written as
\begin{equation}
\label{eq:anesthesia_1.1}
P\cdot \mbox{[ED]}_{50}=\mbox{const.}
\end{equation}
In other words: The more soluble an anesthetic is in a biomembrane (which has the features of an oil), the higher is its anesthetic potency, i.e., the less of the drug is required in order to induce anesthesia. This finding is called the Meyer-Overton correlation and has been known since 1901 \cite[]{Overton1901, Overton1991}. It is remarkable that this finding is absolutely independent of the molecular structure of the general anesthetic and includes drugs as different at NO$_2$ (laughing gas), xenon (a noble gas) and sevofluorane (C$_4$H$_3$F$_7$O). Thus, more specifically the Meyer-Overton correlation states that at critical dose the molar fraction of all general anesthetics in the biological membrane is always exactly the same. The Meyer-Overton correlation is an experimental finding that did not find an explanation in the original publications by Meyer and Overton. The validity of the correlation has been challenged because there exist membrane-soluble molecules such as cholesterol and long-chain alcohols that are no anesthetics. However, if one assumes that an anesthetic is readily soluble in the liquid phase of the membrane and insoluble in the solid phase, thermodynamics tells us that it will lower the melting temperature of the membrane and the Meyer-Overton correlation will be respected. One arrives at the following law:

\begin{equation}
\label{eq:anesthesia_1.2}
\Delta T_m=-\frac{RT^2}{\Delta H}x_A \;,
\end{equation}
where $\Delta T_m$ is the change in the melting temperature caused by anesthetics, and $x_A$ is the molar fraction of anesthetics in the membrane. In this equation $R$ is the gas constant, $T$ the absolute temperature and $\Delta H$ the heat of melting of the membrane. This law has first been derived by van't Hoff in 1887 \cite[]{vantHoff1887}. It is known as the freezing-point depression law. Kharakoz showed that the melting-point depression law is correct for the alcohols from ethanol to decanol \cite[]{Kharakoz2001} but it is also found to be correct for many general and local anesthetics \cite[]{Heimburg2007c, Graesboll2014}. Fig. \ref{graesboll2014} illustrates this for the general anesthetic octanol and and the local anesthetic lidocaine. Both calculated and measure melting profiles in an artificial membrane are shown. Those membrane-soluble drugs that do not cause anesthesia have been show to not cause melting-point depression (discussed in \cite{Graesboll2014}).

Thus, it is fair to say that all anesthetics are anti-freeze agents for biomembranes and obey eq. (\ref{eq:anesthesia_1.2}). For general anesthetics, at the critical dose of anesthetics the shift of the melting transition towards lower temperature is always the same independent of the chemical nature of anesthetics. For local anesthetics, a critical dose is not defined, but the correlation between melting-point depression and partition coefficient is the same as for general anesthetics \cite[]{Graesboll2014}. This effect is of purely physical nature. It finds its explanation in the entropy of ideal mixing of the anesthetic in the liquid membrane. Thus, this finding is completely incompatible with the notion of a specific binding of anesthetic drugs to a molecular receptor which would require non-ideal mixing. All anesthetics have the same effect at the same membrane concentration as xenon, which is a noble gas and cannot bind specifically. For tadpoles $\Delta T_m \approx -0.6$ degrees as measured by their immobilization (i.e., when they sink to the ground). In humans, the change in transition temperature is probably somewhat higher because the measure for the critical dose is different. It is probably of order $\Delta T_m \approx -1$ degree.

Melting-point depression provides a physical explanation for the Meyer-Overton correlation. However, it would remain just another correlation if it did not have predictive power. If the change in the biological melting temperature is the cause of anesthesia, it must be required that anesthesia can be reversed by other changes in the physical conditions that lead to an increase in phase transition temperature. For instance, it is known that melting transitions both in artificial lipid membranes and in biomembranes increase upon hydrostatic pressure by approximately 1$^\circ$ per 40 bar \cite[]{Ebel2001, Heimburg2007c, Muzic2019}. For a shift of +0.6\$\^{}\textbackslash{}circ\$, a pressure of about 25 bars is required in synthetic lipids. At approximately this pressure anesthesia should be reversed if the above correlations are relevant. In fact, it is well-known that anesthesia can be reversed by pressures around 50 bars in tadpoles and other animals that tolerate high pressure \cite[]{Johnson1950, Johnson1970, Miller1973}. This can be taken as evidence that the change in melting temperature induced by anesthetics is physiologically meaningful. Below we will also discuss the interference of anesthesia with inflammation (low pH) and lithium.


\section{Channels in lipid membranes}
\label{channelsinlipidmembranes}

\begin{figure}[htb]
\centering
\includegraphics[width=8.5cm]{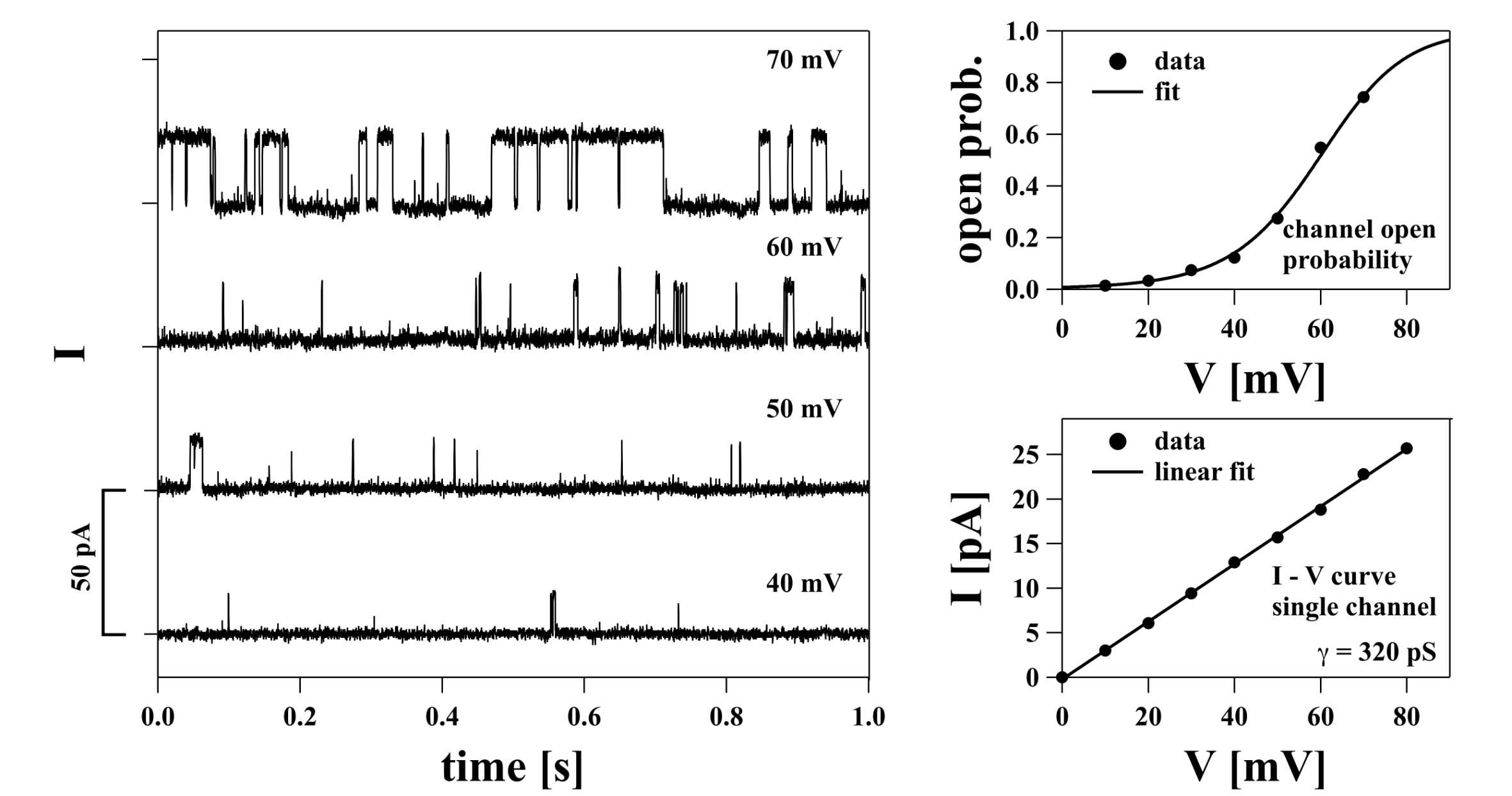}
\caption{\label{blicher2013}Left: Ion channels in a DMPC:DLPC=10:1 membrane as a function of voltage. Bottom, right: The single channel current is proportional to voltage with a single channel conductance of $\gamma=320$ pS. Top, right: The open probability of the pore is sigmoidal with a threshold voltage of about 40 pA. This is very similar to many K$^+$-channels. Adapted from \cite{Blicher2013}.}
\end{figure}

The soliton theory is also consistent with the emergence of ion channel conductance events if they are considered from a thermodynamic perspective. It has been shown that lipid membranes in the complete absence of ion channel proteins can display ion channel-like conduction events whose appearance that is indistinguishable from channels believed to be induced by proteins (see Fig. \ref{blicher2013}, left). This happens in particular close to the melting transitions \cite[]{Antonov1980, Antonov2005, Blicher2009, Wunderlich2009, Laub2011, Heimburg2010} that are the central requirement for the creation of solitons in nerves. These events are influenced by changes in temperature, calcium concentration, membrane tension, and voltage (see \cite{Heimburg2010} and references therein, \cite{Blicher2013, Zecchi2021}) and pH \cite[]{Kaufmann1983a, Kaufmann1983b}. In particular, they are voltage-gated, i.e., the open probability of the channels increases with increasing voltage (see Fig. \ref{Blicher2013}, right) --- a property that was believed to be indicative of voltage-gated protein ion channels (discussed in \cite[]{Zecchi2021}).

The increased permeability of membranes and the emergence of lipid channels close to melting transitions finds its explanation in the greater compressibility of membranes close to phase transitions \cite[]{Nagle1978b, Heimburg2010} and in the thermal fluctuations of the membrane. The open probability is a function of the membrane compressibility, and the channel life-times reflect the fluctuation time-scale. Any change in a variable such as voltage, temperature, pH, pressure, anesthetics concentration etc. will have the potential to increase the open-probability or closed-probability of lipid channels in the complete absence of any molecular receptor that could be blocked. Lipid channels are controlled by the properties of the macroscopic thermodynamics of the membrane. Although they are of nano\-scopic dimensions \cite{Boeckmann2008, Cordeiro2018, Kirsch2019}, their features are not properties of a single macromolecule. They respond to the same changes in the thermodynamic variables as the excitability of the nerve membrane, and anesthesia in the soliton theory. Thus, any stimulus that can excite the nerve membrane will also lead to the formation of lipid pores, and any change that inhibits pulse excitation will seemingly ``block'' lipid channels. E.g., if a membrane is slightly above its melting transition, cooling will open channels, while heating will ``block'' them \cite[]{Blicher2009} in a manner that is very similar to the action of TRPM8 channels \cite[]{Mosgaard2013b}. Similarly, if the experimental temperature is at the heat capacity maximum, anesthetics lower the melting temperature and ``block'' channels \cite[]{Blicher2009}. The term ``block'' is in quotation marks because there is no macromolecule that is blocked but rather a change in the thermodynamic properties of the overall membrane that makes pore formation less likely. Thus, the disappearance of channel events with the addition of a drug does not prove that a receptor is blocked by specific binding. However, the belief in this invalid argument is required in the study of channel proteins, where the possibility of pore formation in the lipid membrane is generally denied.


\section{Possible implications: \\Why should medicine care?}
\label{possibleimplications:whyshouldmedicinecare}

As we have seen, the existence of phase transitions in the nerve membrane slightly below physiological temperature is an essential requirement in the soliton theory. In the following we will address the possibility that the exact value of the transition temperature is important and that deviations lead to physiological malfunction.

\begin{figure}[htbp]
\centering
\includegraphics[width=7cm]{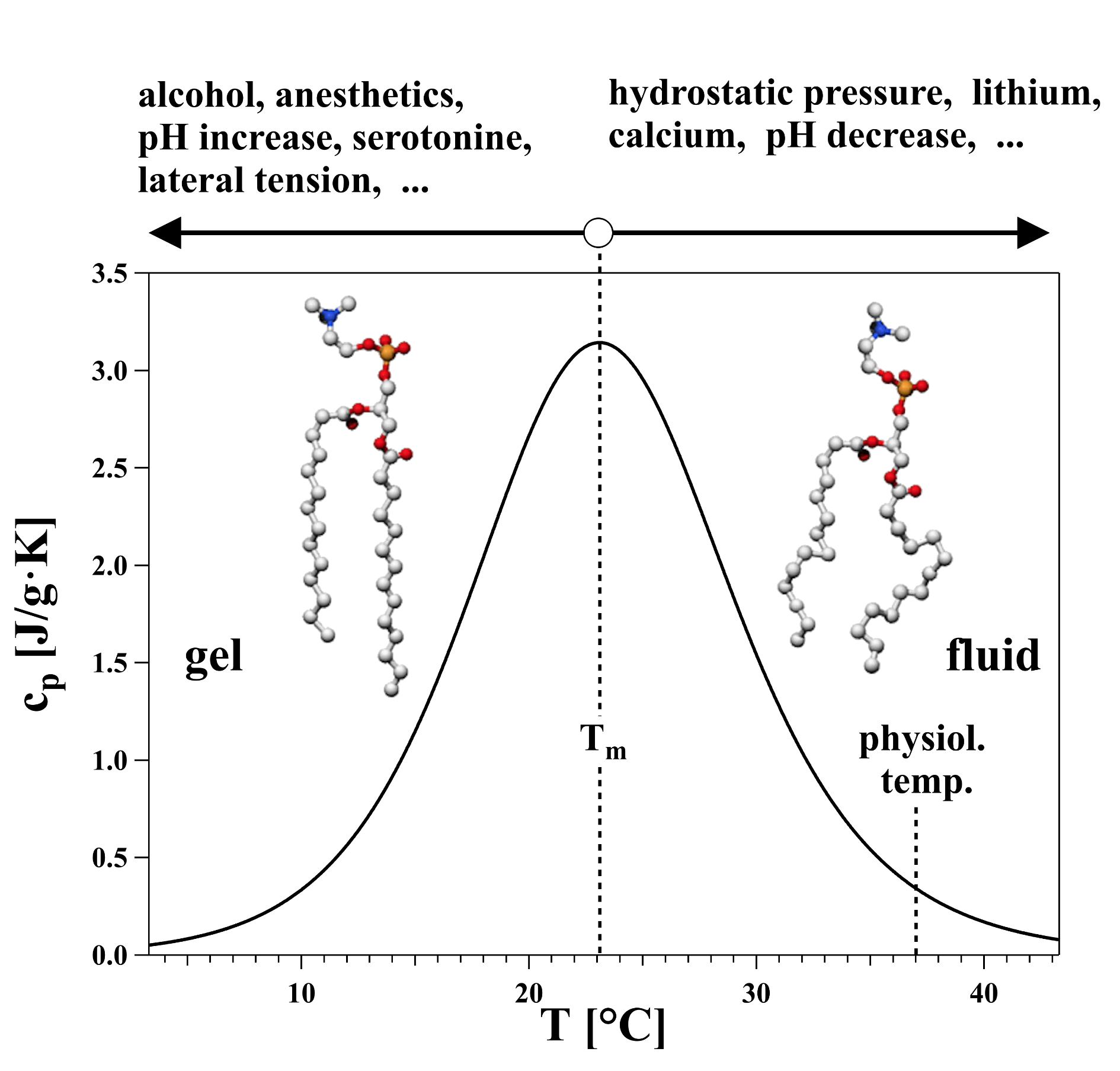}
\caption{Melting transition in a biomembrane (theoretical) with a transition midpoint around 23$^\circ$C. The melting point increases with hydrostatic pressure, lithium and calcium concentration and pH decrease, which results in similar changes in the membranes as decreasing the experimental temperature. The melting point decreases upon increase of the concentrations of anesthetics and alcohol, of serotonin, increase of lateral tension, which results in similar changes as increasing the experimental temperature. The conformations of lipids in the gel and the fluid state are shown below and above the transition.}
\label{biomembranetransition}
\end{figure}

The phase transition temperature responds to changes in all thermodynamic variables (see also \cite{Mussel2019, Mussel2019b}). Some changes, e.g., lowered temperature, increase in lithium and calcium concentrations, pressure, etc, make the membrane more solid and render a nerve more excitable. Other changes (higher temperature, mechanical stretch in nerves, presence of alcohol or anesthetics, etc) make the membranes more fluid and the nerves less excitable. This is schematically shown in Fig. \ref{biomembranetransition}. If two variables are changed at the same time, one change can counteract the other , e.g., anesthesia can be counteracted by pressure or inflammation, essential tremor (hyper-excitability) can be counteracted by alcohol, lithium can also be counteracted by alcohol, and without alcohol but in presence of lithium one can generate tremor. Such phenomena remain obscure in the Hodgkin-Huxley model because the different variables are not trivially coupled. Most thermodynamic variables are not considered at all.

In the soliton theory, the excitability of nerve membranes is determined solely by the temperature of the phase transition in the nerve membrane. The closer the transition is to physiological temperature, the more excitable it is because the free energy required to move the membrane through its transition \cite[]{Wang2018} becomes smaller. Increasing the melting temperature or making the membrane more solid will increase the nerve excitability, and decreasing the phase transition temperature or making the membrane more liquid will reduce excitability. The transition temperature is the switch that responds to many changes in the thermodynamic variables. The possibility to balance changes in several of those variables that occur at the same time is the main strength of the thermodynamic couplings described above. It is the basis for the compensatory treatments suggested below.


\subsection{Anesthesia and Drugs}
\label{anesthesiaanddrugs}

In the soliton theory, anesthesia is a consequence of reducing the melting temperature of a biological membrane (see section \ref{anesthesia}). This is true for both general and local anesthetics \cite{Graesboll2014}. An increase of the anesthetic dose affects the stimulation threshold for an action potential, which becomes higher in agreement with experience \cite{Moldovan2014, Wang2018}. Since the transition temperature depends on all intensive variables, all variables will have some effect on anesthesia. Thus, anesthesia can be counteracted by hydrostatic pressure \cite[]{Johnson1950, Johnson1970, Miller1973}, by decreasing pH \cite[]{Heimburg2007c} or by any other change that increases the transition temperature.


\subsubsection{Finding a good anesthetic}
\label{findingagoodanesthetic}

The above should make it possible to identify potentially good anesthetics by simple laboratory tests. Required is the determination of the oil-water partition coefficient of the drug, and the determination of its influence on the melting temperature of a model membrane. The drug should be insoluble in the solid phase of the membrane and ideally soluble in the liquid phase of the membrane. If these criteria are both met, melting-point depression according to eq. (\ref{eq:anesthesia_1.2}) will result. Substances such as cholesterol or long-chain alcohols that are soluble in both phases are not anesthetics and will not cause melting-point depression \cite[]{Kaminoh1992, Graesboll2014}. Evidently, substances which are most effective in lowering the melting point and follow van't Hoff's melting-point depression law are likely to be the best anesthetics.


\subsubsection{Alcoholism}
\label{alcoholism}

As discussed above, bacteria adapt their lipid composition in response to permanent changes in pressure, temperature or drug concentrations. Since ethanol is an anesthetic (as are all alcohols up to decanol), the permanent presence of ethanol will lead to an adaptation of the lipid composition. Under chronic administration of ethanol, cats adapt the lipids in their livers such that the fraction of polyunsaturated lipids decreases \cite[]{Pawlosky1995}. Likewise, chronic alcoholism in humans decrease decreases the concentration of polyunsaturated lipids \cite[]{Pita1997}. This change in lipid composition will increase the transition temperature of membranes in the absence of alcohol but the constant presence of alcohol will compensate for them. It seems reasonable to speculate that cell membrane function of alcoholics in the presence of alcohol is similar to that of non-alcoholics in its absence. The prediction is that, in the sudden absence of alcohol, the nerves are more excitable than usual due to the increase in the transition temperature of the membranes induced by the adaptation process. According to the soliton theory, this will cause hyperexcitability and possibly pain and tremor. Further, during an operation a sober alcoholic is predicted to require a higher dose of anesthetics than a non-drinker. This is a known fact in clinical praxis \cite[]{Han1969, Fassoulaki1993}. An alcoholic in the presence of his normal dose of alcohol is predicted to require an anesthetic dose equal to that of a non-drinker in the absence of alcohol. In contrast, if a non-drinker happens to be very drunk on a particular occasion, the membrane composition has not adapted the lipid composition and the anesthetic dose will therefore be smaller than normal. The connection of adaptation of membranes to alcoholism has also recently discussed by Schneider \cite{Schneider2021} in a similar context.


\subsubsection{Neurotransmitters}
\label{neurotransmitters}

Some neutrotransmitters such as serotonin also lower the melting temperature of membranes \cite[]{Seeger2007}. Thus, serotonin will exhibit an anesthetizing effect and, like other general anesthetics like laughing gas, ethanol and diethylether, cause euphoria at low doses. The anesthetic action of some neurotransmitters has been discussed previously \cite[]{Cantor2003}. 


\subsection{Neurological disorders}
\label{neurologicaldisorders}


\subsubsection{Essential tremor and alcohol}
\label{essentialtremorandalcohol}

Essential tremor (ET) is among the most common movement disorders. It consists of tremor of the upper limbs, for reviews see \cite{Elble2013, Boutin2015}. Essential tremor is often diagnosed when other origins of tremor can be ruled out \cite{Espay2017}. Let us tentatively assume that ET is related to an increased excitability of nerves. In the thermodynamic picture this suggests an increase of the membrane melting temperature in the nerve membranes. Mechanisms that lower melting temperature should tend to suppress tremor; those that raise melting temperature should cause tremor. As described above, anesthetics lower the melting transition due to the freezing-point depression law and therefore should reduce tremor. Interestingly, it has been \linebreak known for some time that alcohol suppresses ET \cite{Growdon1975}. This has been confirmed by many studies \cite{Koller1984, Koller1994, Zeuner2003, Klebe2005}. Alcohol sensitivity is a secondary criteria for the diagnosis of ET \cite{Knudsen2011}. This is in agreement with the prediction from thermodynamics.


\subsubsection{Bipolar disorder, lithium, tremor and alcoholism}
\label{bipolardisorderlithiumtremorandalcoholism}

Lithium is a commonly used drug in the treatment of bipolar disorder. It was introduced in clinical praxis by J. F. Cade and especially by M. Schou \cite{Cade1949, Schou1954, Schou1958, Schou2001}. It is administered in high doses over several weeks before it is effective. Therapeutic serum levels around 0.8 - 2 mM \cite[]{Sproule2002} speak against any specific binding to a receptor because specific binding would go along with much lower critical doses. Its working mechanism is yet unknown.
It has been shown that, in contrast to other monovalent cations such as sodium and potassium, lithium increases phase transitions in charged lipid membranes due to its binding to charged groups \cite[]{Hauser1981, Hauser1983}. Since about 10--40\% of the lipids in biological membranes are charged \cite[]{Heimburg2007a}, lithium has a profound effect that is similar to that of calcium \cite[]{Binder2002}. Thus, in the soliton theory, lithium render the nerve membrane more excitable. Assuming again that tremor is related to an over-excitability of the nerves, the prediction is that lithium may cause tremor. There is some clinical support for this expectation. Tremor is a side effect of lithium administration \cite[]{Stahl2021}. It has been argued that lithium-induced tremor is a subtype of essential tremor \cite{Hallett1986, Baek2014}. Further, it has been shown that lipid compositions changes with the chronic administration of lithium, and membranes become more fluid by adapting their unsaturated lipid composition \cite[]{LopezCorcuera1988}. This further supports the notion that membranes adapt to permanent changes in the thermodynamic variables. Since lithium increases the transition temperature, it should tend to counteract the effects of alcohol . It has in fact been found that alcohol and lithium have opposing action in bipolar patients, and chronic alcoholism increases bipolar symptoms \cite{Nascimento2015}. The anesthetic ketamine is often used to treat depressions while lithium is used to attenuate mania \cite[]{Stahl2021}. Thus, anesthetics and lithium seem to have opposing effects.


\subsection{Inflammation}
\label{inflammation}

Inflamed tissue is more difficult to anesthetize \cite[]{Harris1964, Heimburg2007c}. This is well-known in dentistry. Neighboring teeth can react differently to the administration of anesthetics if one of them shows inflammation of the root. This effect has been attributed to an increased firing rate of the nerves in the inflamed tissue. This may be related to the pH-shifts caused by inflammation \cite[]{Rood1982}.
Melting transitions in biological membranes can be shifted by changing pH. This has been found for both charged artificial members and biological membranes \cite[]{Trauble1976, Muzic2019, Fedosejevs2022}. Proton binding to charged lipids will lead to higher melting temperatures due to the reduction of the electrostatic surface potential. In native \emph{E.coli} membranes, changing the pH from 9 to 5 increases the transition by 7$^\circ$ \cite[]{Muzic2019}. Thus, lowering pH is predicted to counteract anesthesia \cite[]{Heimburg2007c}. This has consequences for inflamed tissue. It is known that inflammation leads to a decrease of pH by about half a pH-unit on average (depending on the severity of the inflammation, \cite{Punnia-Moorthy1987, Gautam2010}). Thus, inflamed tissue will display a higher transition temperature, will be more excitable in agreement with experiment (thus potentially causing pain), and thus be more difficult to anesthetize. \\


\subsection{Stretching of nerves}
\label{stretchingofnerves}

It has been known for more then 140 years that nerves can be excited mechanically \cite[]{Tigerstedt1880}. Further, mechanical changes in neurons have been observed in connection with the action potential \cite[]{Iwasa1980a, Iwasa1980b, Tasaki1989, Kim2007, GonzalezPerez2016, Ling2020}. Recently, \cite{Ucar2021} showed that synapses react to stretching which implies a mechanosensitivity of synapses. Furthermore, nerves contract upon excitation \cite[]{Wilke1912b}. Tasaki and collaborators \cite{Tasaki1989} found a contraction of garfish olfactory nerve while the action potential was traveling. This implies that during the action potential a neuron is under increased tension when the ends are fixed. This is predicted by the soliton theory, in which nerves will contract during the time period in which the action potential travels because the pulse consist of a region of gel membrane with smaller area density. In general, stretching of nerves is expected to make them less excitable since lateral tension in a membrane leads to a shift of the phase transition towards lower temperatures \cite[]{Wang2018}. Experiments show that stretching causes changes in nerve function. As early as 1935 \cite{Hayasi1935} showed that the chronaxie of motor nerve stimulation changes with stretching. The effect of stretching on reflexes has been studied extensively \cite[]{Budini2016}. \cite{Guissard2001, Guissard2006} showed that motorneuron excitability decreased during fast muscle stretch. This inhibition had been tested in the soleus muscle by recording the Hoffmann reflex (H reflex) and the tendon reflex (T reflex) which were both found to be reduced during fast muscle stretching. Compound action potentials decrease in amplitude during mechanical stretch in hamster sciatic nerve \cite[]{Stecker2011} and rat nerves \cite[]{Ochs2000}. The action potentials of rabbit tibial and peripheral nerves also decreased in amplitude upon stretching \cite[]{Wall1992, Kwan2009}. This is in agreement with predictions from the thermodynamic theory in which increased mechanical tension reduces excitability. For a more detailed discussion of the effect of stretching in the context of the soliton theory see \cite{Heimburg2022a}.


\subsection{Focused ultrasound neurostimulation}
\label{focusedultrasoundneurostimulation}

Focused ultrasound neurostimulation of the brain is an emerging area of clinical interest \cite[]{Tufail2011, Tyler2012, Mueller2014, Rezayat2016, Tyler2018}. It has been found that focussed ultrasound can stimulate nerves from a distance, i.e., it is a non-invasive technique. It has high spatial resolution and the ability to reach deep brain targets \cite[]{Sanguinetti2020}. \cite{Kamimura2016} showed how one selectively can stimulate brain regions in mice by focussing ultrasound such that they contract the left or the right leg, or the tail.

It is not understood how ultrasound causes nerve stimulation, but in several studies have considered the possibility that the soliton theory offers an explanation of the mechanism \cite{Tufail2011, Mueller2014, Jerusalem2019, Kamimura2020}. This would indeed make sense since the nerve pulse has a mechanical component.
Therefore, it would be interesting to compare the typical frequencies and wavelength of pulse ultrasound neuromodulation with the length and time scale of nerve pulses.
The time scale of the nerve pulse is of the order of 1ms, corresponding to the period of a carrier frequency of $\nu=1$ kHz. When ultrasound is applied to the brain, it propagates either in the aqueous cytosol or in the nerve membranes. The sound velocity in water is 1500 m\slash s, and in a motor neuron it is of order $c=100$ m\slash s \cite{Heimburg2005c}. The wavelength is given by $\lambda=c/\nu$. The wavelength in water and fluid membranes for different frequencies is given in Table \ref{table2}. 

\begin{table}[hb]
\begin{center}
\begin{tabular}{ |c|c|c| } 
 \hline
frequency & wave length &  wave length \\ 
  & water & nerve membrane \\ 
\hline\hline
 1KHz & 1.5 m & 10 cm \\ 
 100 KHz& 15 mm & 1 mm \\ 
 1 MHz& 1.5 mm & 0.1 mm \\ 
 \hline
\end{tabular}
\end{center}
\caption{\label{table2}Wavelength of ultrasound in water (assuming a sound velocity of 1500 m/s) and membranes (assuming a 2D sound velocity of 100 m/s \cite{Heimburg2005c} ) at three different frequencies. Compare the wave lengths with the length of the nerve pulse given in Table \ref{table1}.}
\end{table}

Thus, the wavelength of sound in water in the frequency range from 100 kHz-1MHz is between 15 and 1.5 mm. An action potential in a non-myelinated nerve displays pulse-lengths of a few mm or cm (see table \ref{table1}). In in-vivo experiment, the stimulation frequency is rather between 250 KHz-7 MHz (see tables in \cite{Tufail2011, Ye2016}). The ultrasound is often pulsed with pulse frequencies in the range of 1--3 kHz (e.g., \cite{Tufail2011, Kamimura2016}), which is close to the characteristic frequency of the nerve pulse with T=1ms.

Summarizing, the wavelength in water of the typical ultrasound frequencies used in in-vivo experiments is of similar order as the length of the nerve pulses in non-myelinated nerves, and the period of the repetition frequency of pulsed ultrasound is similar to the characteristic frequency of nerve pulses. It remains to be investigated whether these similarities are meaningful for focused ultrasound neurostimulation. However, experiments have evolved to just favor these frequencies, which might not be accidental.

\section{Conclusion}
\label{conclusion}

We discuss the advantages that a macroscopic thermodynamic view may have for the understanding of nerve and cell function. In contrast to the Hodgkin-Huxley model, the soliton theory is deeply rooted in thermodynamics. Its central ingredient is the phase transition below body temperature that responds to changes in all thermodynamic variables. The transition acts like a switch that can be shifted downwards or upwards in temperature. The membrane as a whole responds to changes in voltage, pressure, tension, pH, drugs and the chemical potentials of membrane ingredients. Two variables can have opposing effects on the temperature of the phase transition. Examples are the known pressure reversal of anesthesia and the reversal of anesthesia by inflammation. A perfect anesthetic drug is predicted to obey van't Hoff's freezing-point depression law. We discuss alcoholism in connection with the adaptation of membranes to the permanent presence of alcohol. We suggest that there exist other examples for compensatory action such as the attenuation of essential tremor by alcohol, or the induction of essential tremor by lithium. Further, stretching of nerves will reduce the excitability of nerves, which may be related to the experimentally measured attenuation of H- and T- reflexes. A thermodynamic view may be able to explain why nerves can be excited by ultrasound because ultrasound frequency and the pulse-repetition frequency match the dimensions of the nerve pulse. Thus, the soliton theory suggests a variety of possible correlations that may be of high medical interest. In such a picture, some medical disorders may be due to undesirable changes in the membrane phase transition temperature due to a macroscopic imbalance that can be offset by adjusting other thermodynamic variables. Such couplings are inaccessible if the effects of changes in variables is attributed to single receptor proteins and would remain obscure.

\vspace{0.5cm}

\textbf{Acknowledgments:} I thank Christine P. Lauritzen and Dana T. Kamp for introducing me to essential tremor and the treatment of bipolar disorder, and Andrew D. Jackson for a critical reading of the manuscript. Prof. Matej Daniel from the Czech Technical University in Prague discussed the importance of phase transitions for medicine with me and encouraged the writing of this publication.

\vspace{0.5cm}


\small{
	
}

\end{document}